\documentclass[preprint,12pt]{elsarticle}
\usepackage{graphicx,amssymb,amsmath,color,numcompress}
\journal{Physics Letters B}

\begin{document}

\begin{frontmatter}



\title{Neutral current (anti)neutrino scattering: relativistic mean field and superscaling predictions}


\author[label0]{R. Gonz\'alez-Jim\'enez}
\author[label1,label2]{M.~V.~Ivanov\corref{cor}}
\cortext[cor]{Corresponding author.}
\ead{martin.inrne@gmail.com}
\author[label4]{M.~B.~Barbaro}
\author[label0]{J.~A. Caballero}
\author[label1]{J.~M.~Udias}

\address[label0]{Departamento de F\'{\i}sica At\'{o}mica, Molecular y Nuclear, Universidad de Sevilla, 41080 Sevilla, Spain}
\address[label1]{Grupo de F\'{i}sica Nuclear, Departamento de F\'{i}sica At\'omica, Molecular y Nuclear, Facultad de Ciencias F\'{i}sicas, Universidad Complutense de Madrid, CEI Moncloa, Madrid E-28040, Spain}
\address[label2]{Institute for Nuclear Research and Nuclear Energy, Bulgarian Academy of Sciences, Sofia 1784, Bulgaria}
\address[label4]{Dipartimento di Fisica Teorica, Universit\`a di Torino and INFN, Sezione di Torino, Via P. Giuria 1, 10125 Torino, Italy}

\begin{abstract}

We evaluate the neutral current quasi-elastic neutrino cross section within two nuclear models: the SuSA model, based on the superscaling behavior of electron scattering data, and the RMF model, based on relativistic mean field theory. We also estimate the ratio $(\nu p \to \nu p)/(\nu N \to \nu N)$ and compare with the MiniBooNE experimental data, performing a fit of the parameters $M_A$ and $g_A^{(s)}$ within the two models. Finally, we present our predictions for antineutrino scattering.

\end{abstract}


\end{frontmatter}

% \linenumbers

\section{Introduction \label{sec:1}}
The study of neutral current mediated quasi-elastic (NCQE) neutrino-nucleus scattering in the GeV region is a powerful tool for hadronic and nuclear studies. We note that although in the tradition of neutrino experiments the term 'elastic', either neutral-current elastic or charged-current elastic is used for neutrino scattering off free nucleons as well as on nucleons bound on nuclei, in this work we will refer to the latter case with the more precise denomination of quasi-elastic (QE). NCQE can be used, on one hand, to obtain information on the structure of the nucleon, in particular on its strange quark content, on the other it represents a probe of nuclear dynamics complementary to neutrino charged current quasi-elastic (CCQE) scattering and electron scattering. Several theoretical investigations have been devoted to the study of this reaction making use of different nuclear models~\cite{Amaro:2006pr,Leitner:2006sp,Antonov:2007vd,Meucci:2011ce,Benhar:2011wy,Martini:2011wp,Butkevich:2011fu,Ankowski:2012ei,Morfin:2012}.

The MiniBooNE experiment~\cite{AguilarArevalo:2010cx} has recently reported a
high-statistics measurement of the NCQE cross section on mineral oil (CH$_2$)
and of the ratio $(\nu p \to \nu p)/(\nu N \to \nu N)$ between single-proton
and proton+neutron cross sections.
In this letter we compare these measurements with the predictions of two
relativistic nuclear models, the Super-Scaling-Approximation (SuSA) and the
relativistic mean field (RMF) models, which have been previously applied to
the CCQE process~\cite{Amaro:2004bs,Amaro:2011qb}. A detailed description of the two models can be found in
Refs.~\cite{Amaro:2004bs} and~\cite{Caballero:2005sj}.
Here we just recall their main ingredients: the SuSA approach is based on
the assumption that the superscaling function~\cite{DonnellySick} extracted
from quasi-elastic electron scattering data can be implemented in the
neutrino-nucleus cross section, the only differences between the two processes
being related to the elementary reaction and not to the nuclear response;
the RMF model provides a microscopic description of the process, where
final-state interactions (FSI) are taken into account by using the same relativistic
scalar and vector energy-independent potentials considered to describe the
initial bound states.
Both models give an excellent representation of the experimental superscaling
function~\cite{Caballero:2005sj}, in contrast to the relativistic Fermi gas
(RFG), which fails to reproduce the electron scattering data.

It has been shown in Ref.~\cite{Amaro:2011qb} that,
when applied to CCQE reactions, the RMF and SuSA models give similar results,
although some difference arises: both models underestimate the MiniBooNE
data~\cite{AguilarArevalo:2010zc}, but the RMF gives a smaller discrepancy.
It has been suggested by various
authors~\cite{Martini:2009uj,Amaro:2010sd,Nieves:2011yp,Meucci:2011vd} that
the gap between theory and data can be filled by meson-exchange currents,
multinucleon emission or particular treatments of final-state interactions.
If one sticks to a simple nuclear description, such as the RFG model,
presently used in neutrino interaction generators, the experimental increase
in the cross section can be obtained by introducing a nucleon axial mass
$M_A=1.35$~GeV, significantly larger than the standard value $M_A=1.03$~GeV~\cite{Bernard:2001rs}, which simulates the additional nuclear effects not considered
in the RFG.

%\section{Formalism \label{sec:2}}
%

\section{Results and discussion \label{sec:3}}

\begin{figure}[t]
\centering
\includegraphics[height=80mm,angle=270]{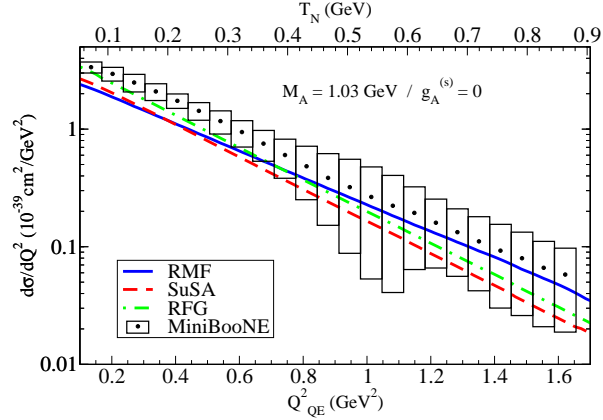}
\caption{NCQE flux-averaged cross section computed using the RMF (solid blue),
 SUSA (dashed red) and RFG (dot-dashed green) models and compared with the
MiniBooNE data~\cite{AguilarArevalo:2010cx}.}\label{fig:CH2-flux-average-models}
\end{figure}

Let us now consider the neutral current (NC) process. In order to compare with MiniBooNE data on
CH$_2$, we evaluate the following differential cross section per nucleon
\begin{multline}
\hspace*{-4.mm}\frac{d\sigma}{dQ^2} =
\frac{1}{7}  C_{\nu p,H}(Q^2)
\frac{d\sigma_{\nu p\to \nu p, H} }{dQ^2}+
\frac{3}{7}  C_{\nu n,C}(Q^2)
\frac{d\sigma_{\nu p\to \nu p, C} }{dQ^2}\\
+\frac{3}{7}  C_{\nu n,C}(Q^2)
\frac{d\sigma_{\nu n\to \nu n, C}}{dQ^2}~,
\end{multline}
which results from three contributions: scattering on free protons,
bound protons in Carbon and bound neutrons in Carbon, each of them weighted
by an efficiency correction function $C_i$ and averaged over the experimental
neutrino flux~\cite{AguilarArevalo:2010cx}.
Results corresponding to the two models mentioned above as well as the RFG are
shown in Fig.~\ref{fig:CH2-flux-average-models} as functions of the
``quasi-elastic'' four-momentum transfer $Q_{QE}$ defined in
\cite{AguilarArevalo:2010cx} or of the outgoing nucleon kinetic energy $T_N$.
The standard value $M_A=1.03$~GeV has been taken for the axial mass,
while the strange quark contribution to the axial form factor at $Q^2=0$,
$g_A^{(s)}$ (or equivalently $\Delta s$), has been set to zero.
For the electric and magnetic strangeness the results of a recent global
analysis of PV electron-proton asymmetry data~\cite{GonzalezJimenez:2011fq}
($\rho_s=0.59$, $\mu_s=-0.02$) have been used.
Note however that the cross section is essentially independent of $\rho_s$, $\mu_s$~\cite{long}.

We note that the SuSA cross section is smaller than the RFG one by about
20\% and the two curves have essentially the same slope in $Q^2$.
On the other hand the RMF result has a softer $Q^2$ behavior, with a smaller
slope. This is at variance with the CCQE case, for which, as shown in
Ref.~\cite{Amaro:2011qb}, SuSA and RMF cross sections are very close to each
other. This result indicates, as expected, that the NC data, for which the
outgoing nucleon is detected, are more sensitive to the different treatment
of final-state interactions than the MiniBooNE CC data, where the ejected
nucleon is not observed.

\begin{figure}[ht]\centering
\includegraphics[height=80mm,angle=270]{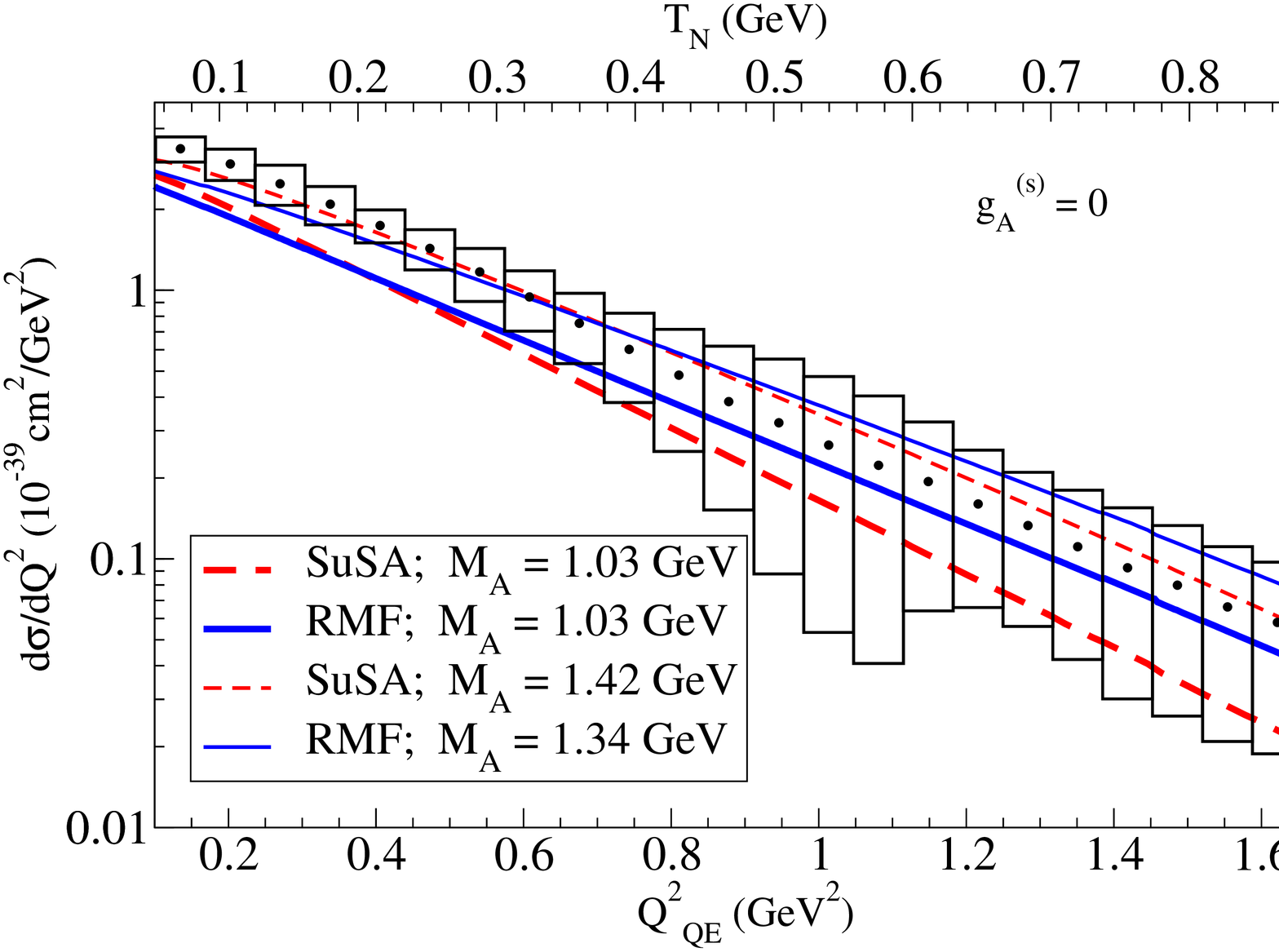}
\includegraphics[height=80mm,angle=270]{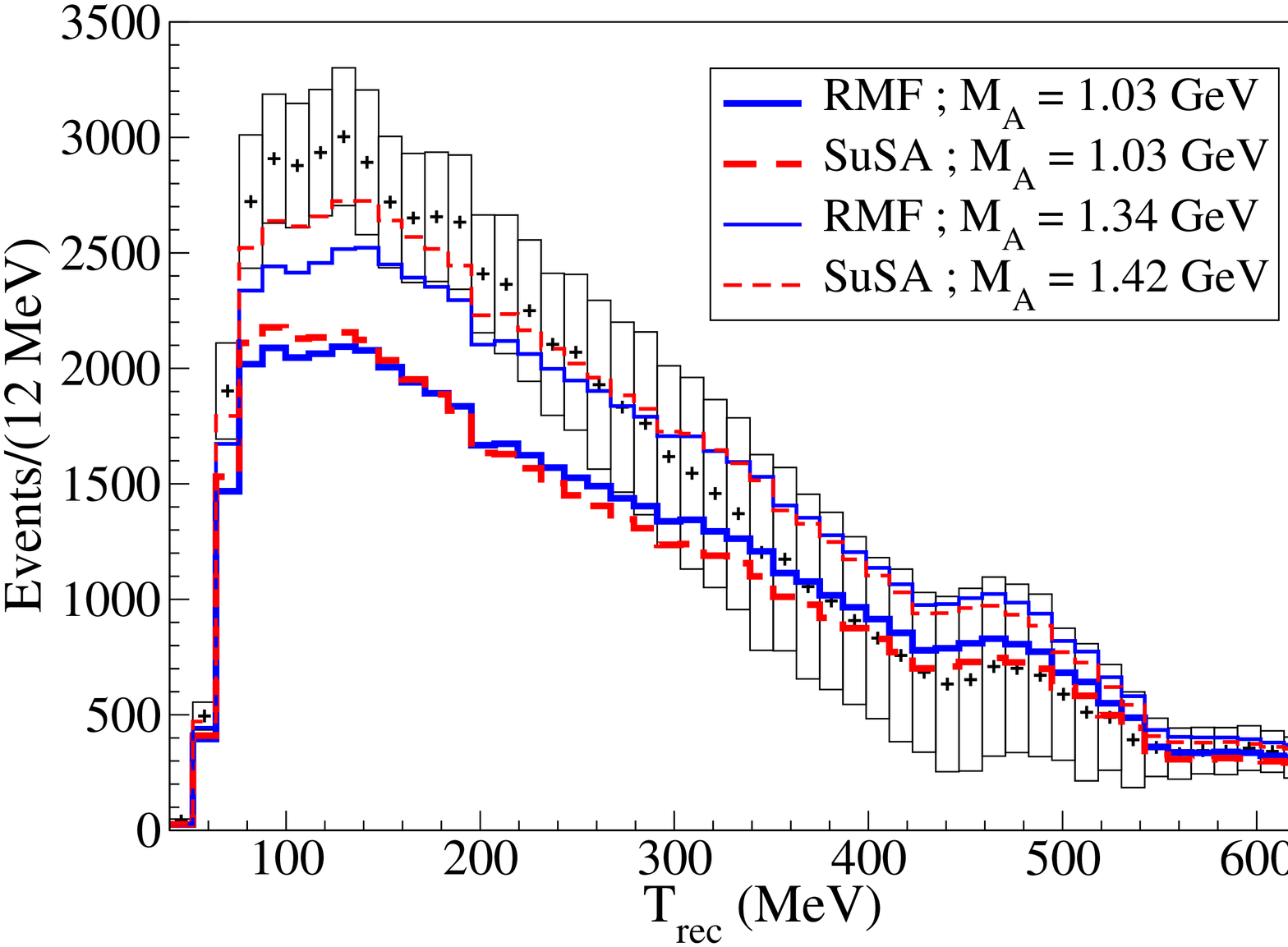}
\caption{NCQE flux-averaged cross section computed within the RMF
(solid blue lines) and SuSA (dashed red lines) models, compared with MiniBooNE
data~\cite{AguilarArevalo:2010cx} as a function of true energy on top panel
and of the reconstructed energy on bottom panel, for different values
of $M_A$ (see text).}
\label{fig:MA_RMF_SuSA}
 \end{figure}

In Fig.~\ref{fig:MA_RMF_SuSA} we illustrate the dependence of the cross
section upon the axial mass $M_A$ at strangeness $g_A^{(s)}=0$.
We compare results with the standard axial mass to the ones obtained with the
value of $M_A$ that provides the best fit to the cross section within
either SuSA or RMF models. We fit the axial mass performing a $\chi^2$ test
using the true energy data from MiniBooNE~\cite{AguilarArevalo:2010cx}
(top panel on Fig.~\ref{fig:MA_RMF_SuSA}) with the following $\chi^2$ definition
\begin{equation}
 \chi^2=\sum_i\left(\frac{\text{CS}^\text{exp}_i - \text{CS}^\text{theo}_i}
{\Delta \text{CS}^\text{exp}_i}\right)^2\,,
\end{equation}
where $\text{CS}^\text{exp}_i$ is the experimental cross section in the i-bin,
$\text{CS}^\text{theo}_i$ is the predicted one and $\Delta \text{CS}^\text{exp}_i$ is
the error in $\text{CS}^\text{exp}_i$.
For $g_A^{(s)}=0$, the 1-$\sigma$ allowed regions of the axial mass for the
two models are
\begin{eqnarray}
\label{eq:MA_RMF}
M_A&=& 1.34\pm 0.06~\text{GeV \ \ for RMF}
\\
\label{eq:MA_SuSA}
M_A&=& 1.42\pm 0.06~\text{GeV \ \ for SuSA},
\end{eqnarray}
corresponding to $\chi^2/DOF=16.5/22$ and $\chi^2/DOF=4.7/22$, respectively.
These have to be compared with $\chi^2/DOF=46.2/22$ (RMF) and
$\chi^2/DOF=45.3/22$ (SuSA) for $M_A=1.03$~GeV.

In Fig.~\ref{fig:MA_RMF_SuSA} the RMF and SuSA results are compared with the MiniBooNE data as functions of the true (top panel) and reconstructed (bottom panel) energies. Whenever a physical quantity is measured there are distortions to the original distribution in the observed quantity. Experimentalists correct the data distribution using unfolding techniques. There is an alternative method, which is to report them in the reconstructed nucleon energy, without applying the unfolding procedure (and corresponding errors). To produce the reconstructed energy results we used the folding procedure detailed in Appendix~B of Ref.~\cite{Perevalov}.
We observe that both models give a reasonably good representation of the data
when the non-standard value of the axial mass is used.
Moreover, we note that the SuSA cross section reproduces quite well the slope
of experimental data, better than RMF one which has a smaller $Q^2$ slope and
falls slightly below the error bars for lowest $Q^2$ data. It is however important to observe that none of the two models is
expected to describe correctly the low-$Q^2$ region, where collective effects
play a dominant role. The values of the axial mass obtained with both models are compatible, within
1-$\sigma$, with the value $M_A=1.35$~GeV employed by the MiniBooNE
collaboration to fit their RFG model to the CCQE data.

It has been known for some time (see, {\it e.g.},
\cite{Barbaro:1996vd,Alberico:1997vh}) that the $g_A^{(s)}$-dependence of the
NCQE neutrino-nucleon cross section is very mild.
This results from a cancellation between the effect of $g_A^{(s)}$ on the
proton and neutron contributions, which are affected differently by the axial
strangeness: by changing $g_A^{(s)}$ from zero to a negative value the proton
cross section gets enhanced while the neutron one is reduced,
so that the net effect on the total cross section is very small.
For that reason, the previous analysis performed to fit the axial mass is
quite independent of the axial strangeness, which we just set to zero.
Thus, once we have obtained the axial mass that fits the neutrino
cross sections, we can look for a different observable that can be more
sensitive to the axial strangeness content of the nucleon.
Variations of the axial strangeness can have a large impact on the ratio
between proton and neutron cross sections. Furthermore, many systematic errors
are canceled~\cite{Ahrens1987} in taking the ratio.

The MiniBooNE experiment cannot measure the $p/n$ ratio because the
$\nu n\to \nu n$ reaction cannot be isolated. However, single-proton events
can be isolated above the Cherenkov threshold, and so it was possible to
construct
two different samples: $\nu N \to \nu N$ (where $N$ is either a proton or a
neutron) with the standard NCQE cuts and a $\nu p \to \nu p$
NCQE proton-enriched sample for which two additional cuts were applied.
The ratio $(\nu p \to \nu p)/(\nu N \to \nu N)$ was reported in
Ref.~\cite{AguilarArevalo:2010cx} as a function of the reconstructed nucleon
kinetic energy $T_{\rm rec}$ from $350$ to $800$~MeV.

We now compare the predictions of our models with the experimental ratio,
using the cross section folding procedure described in \cite{Perevalov}.
Following this procedure we convert our NCQE `true energy' cross section into
NCQE reconstructed energy distributions for the numerator and
denominator samples, separately, and finally we take the ratio.
As expected, the ratio, unlike the cross section, is sensitive to axial
strangeness. We now set the axial mass to the values
(\ref{eq:MA_RMF},\ref{eq:MA_SuSA}) previously found from the best fit of the
cross sections at $g_A^{(s)}=0$, and perform a $\chi^2$ fit to the axial
strangeness parameter. The 1-$\sigma$ allowed regions turn out to be
\begin{eqnarray}
\label{eq:gas_RMF}
g_A^{(s)} &=& +0.04\pm0.28 \text{ \ \ for RMF}
\\
\label{eq:gas_SuSA}
g_A^{(s)} &=& -0.06\pm0.31 \text{ \ \ for SuSA},
\end{eqnarray}
corresponding to
$\chi^2/DOF=33.6/29$ and $\chi^2/DOF=31.3/29$, respectively.

\begin{figure}[ht]\centering
         \includegraphics[width=75mm,angle=0]{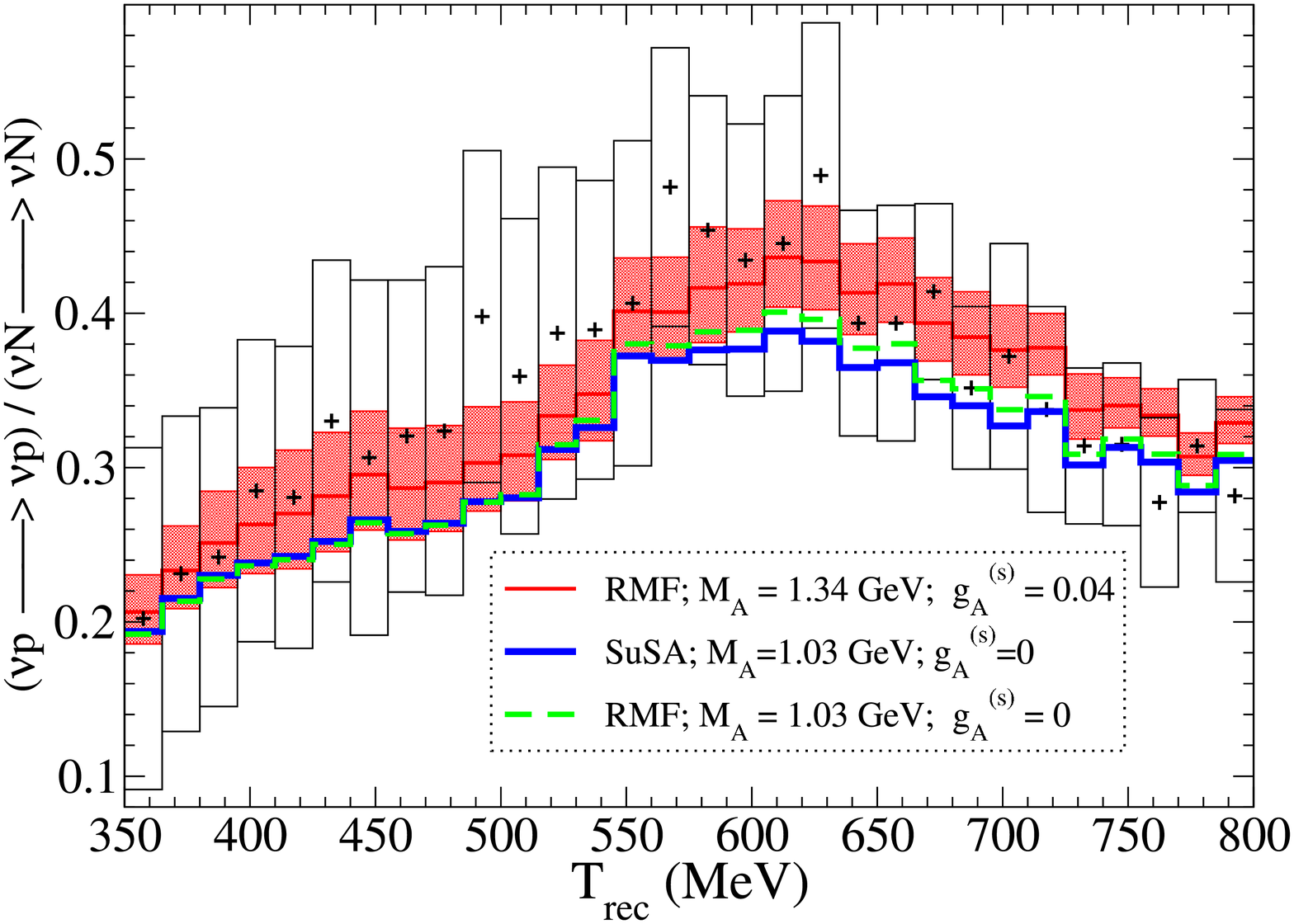}
         \includegraphics[width=75mm,angle=0]{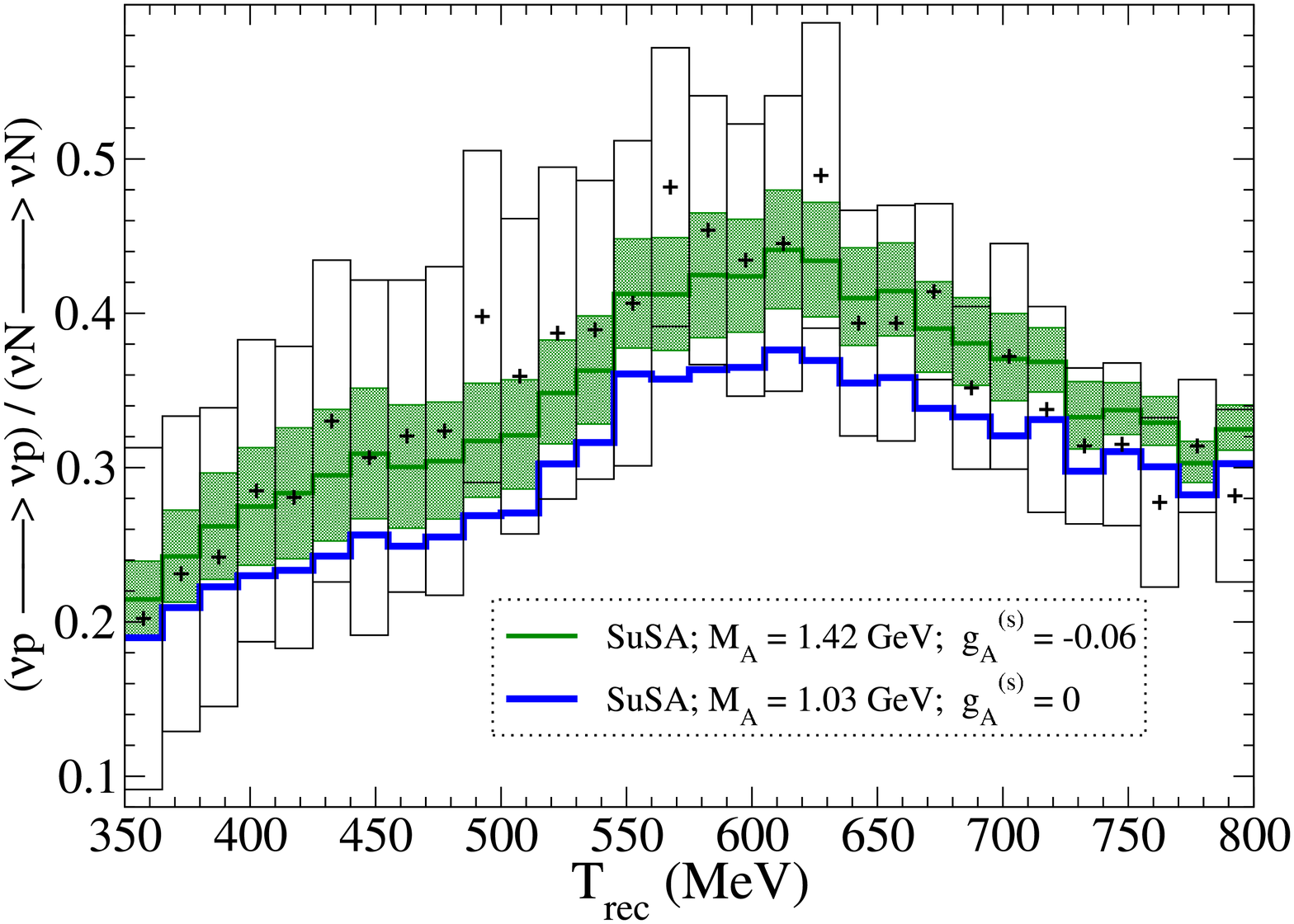}
     \caption{Ratio $(\nu p \to \nu p)/(\nu N \to \nu N)$ computed within RMF and SuSA models. Shadowed
areas represent the 1-$\sigma$ region allowed for $g_A^{(s)}$ (see text). The ratio computed with the
best-$g_A^{(s)}$ is presented as well as those obtained with the standard axial mass and no strangeness.
Data from Ref.~\cite{AguilarArevalo:2010cx}.}
     \label{fig:ratio}
 \end{figure}

In Fig.~\ref{fig:ratio} we present the ratio computed by using the above sets
of parameters as well as with the standard axial mass and no strangeness,
as reference.
From the comparison between the two
curves in the top panel having standard parameters $M_A=1.03$ and $g_A^{(s)}=0$
(green and blue lines), it appears that the dependence upon the nuclear model
is essentially canceled in the ratio, confirming that this is a good
observable for determining the axial strangeness content of the nucleon.
Within the error bars, the values of $g_A^{(s)}$ obtained are compatible with
the ones of the previous analysis. Of course before drawing definitive
conclusions on the allowed value of $g_A^{(s)}$, an extended analysis of the
nuclear effects that are being effectively incorporated in the increased value
of $M_A$ should be performed.
However, it is worth mentioning that the ratio shown in Fig.~\ref{fig:ratio}
shows little sensitivity to a possible np charge-exchange due to FSI. For instance a 20\% of charge-exchange would not affect the results
displayed in Fig.~\ref{fig:ratio} by more than a few percent, for any
reasonable value of $g_A^{(s)}$.

\begin{figure}[t]\centering
         \includegraphics[height=80mm,angle=270]{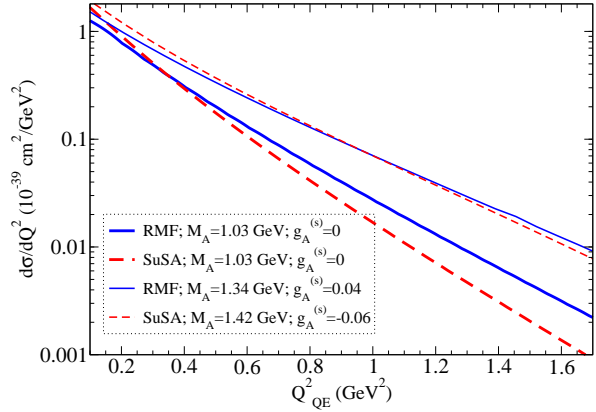}
     \caption{NCQE antineutrino cross section computed using RMF and SuSA  models for different values of $g_A^{(s)}$ and $M_A$. We employed the antineutrino flux prediction for MiniBooNE given in Ref.~\cite{flux}.}
     \label{fig:cs-antineutrino}
 \end{figure}

Before concluding we show our predictions for the NC antineutrino cross
sections. In this case cross sections are slightly more sensitive to the axial
strange content than neutrino ones, especially at high $Q^2$~\cite{Amaro:2006pr,long}.
This is illustrated in Fig.~\ref{fig:cs-antineutrino}, where we display the flux-averaged NCQE antineutrino
cross sections as a function of $Q^2_{QE}$ as obtained within the RMF and SuSA
models. We use the set of parameters (\ref{eq:MA_RMF},\ref{eq:MA_SuSA},\ref{eq:gas_RMF},\ref{eq:gas_SuSA}),
favoured by our analysis of MiniBooNE NCQE neutrino cross section and ratio.
As a reference, results for the standard axial mass and axial strangeness
equal to zero are also presented.

\section{Conclusion \label{sec:4}}

Summarizing, we have presented predictions for MiniBooNE NCQE neutrino cross sections with two nuclear
models, SuSA and RMF. As it was the case of CCQE data, with the standard value of the axial mass,
these models underpredict the cross section data. We have used the axial mass as an effective parameter
to incorporate nuclear effects not used in the models, such as multi-nucleon knockout.
In doing this, we could restore agreement of the models to the data, with axial mass value consistent
with the ones obtained in a similar fit to CCQE data by the MiniBooNE collaboration.
The nuclear models tuned this way can be employed to analyze NCQE cross section ratios as the ones
obtained in the MiniBooNE experiment. We remind the reader that the present calculations are based on
one-particle-one-hole assumptions and meson-exchange currents have not been considered. Provided that the models, as done here, are tuned to reproduce
the total cross section data, our analysis shows that the ratio does indeed show a strong sensitivity to
the axial strangeness content of the nucleon, while being highly model independent.
This shows the relevance of having extended, good statistics data, eventually including antineutrino
cross sections under similar conditions, which may help to disentangle the properties of the
 neutrino-nucleon and neutrino-nucleus interactions, of paramount importance for neutrino-oscillation
experiments.

\section*{Acknowledgements}

This work was partially supported by Spanish DGI and FEDER funds (FIS2011-28738-C02-01), by the Junta de Andalucia, by the Spanish Consolider-Ingenio 2000 program CPAN (CSD2007-00042), by the Campus of Excellence of Moncloa project (Madrid) and Andalucia Tech, partly by the INFN-MICINN collaboration agreement (AIC-D-2011-0704), as well as the Bulgarian National Science Fund under contracts No. DO-02-285 and DID-02/16-17.12.2009. M.V.I. is grateful for the warm hospitality given by the UCM and for financial support during his stay there from the Centro Nacional de F\'isica de Part\'iculas, Astropart\'iculas y Nuclear (CPAN) of Spain and FPA2010-17142. R.G.J. acknowledges support from the Ministerio de Educaci{\'o}n (Spain). The authors would like to thank T.W. Donnelly for careful reading of the manuscript and helpful discussions.

\end{document}